# Acoustic scattering-extinction cross section and the acoustic force of electrostatic type

Ion Simaciu[1,a], Zoltan Borsos[1,b], Gheorghe Dumitrescu[2], Glauber T. Silva[3] and Tiberiu Bărbat[4]

[1] Petroleum-Gas University of Ploieşti, Ploieşti 100680, Romania

[2] High School Toma N. Socolescu, Ploieşti, Romania

[3] Physical Acoustics Group, Instituto de Física, Universidade Federal de Alagoas, Maceió, AL 57072-970, Brazil

[4] Virtual-Ing, Bucuresti, Romania

## Abstract

*The analysis of the secondary Bjerknes force between two bubbles suggests that this force is analogous to the electrostatic forces. The same analogy is suggested by the existence of a scattering cross section of an acoustic wave on a bubble. Our paper brings new arguments in support of this analogy. The study which we perform is dedicated to the acoustic force and to the scattering cross section at resonance in order to highlight their angular frequency independence of the inductor wave. Also, our study reveals that the angular frequency and the amplitude of the induction pressure wave are not related. Highlighting this analogy will allow us a better understanding of the electrostatic interaction if the electron is modeled as an oscillating bubble in the vacuum.*

## 1. Introduction

The aim of this paper is to bring new arguments for the analogy between the electromagnetic world, i.e. the world in which systems and phenomena interact and correlate with electromagnetic waves, and the acoustic world, i.e. the world in which systems and phenomena interact and correlate with acoustic waves. The paper highlight that the acoustic forces of electrostatic type and the cross section of the acoustic interaction are not dependent of the angular frequency of the waves that induce the oscillations of the bubbles.

The analogy between secondary Bjerknes force and the electrostatic force was notified since the early theoretical and experimental study of these forces [1-4]. A general analysis of this analogy was made by Zavtrak [5, 6].

The analysis of the secondary Bjerknes forces between two bubbles reveals that these forces imply the scattering of the inductor acoustic wave, i.e. the forcing waves. The two elastic bubbles absorb energy from the forcing wave and oscillate in volume. These spherical oscillations produce the spherical wave pressure whose intensity decreases as $1/r^2$. The intensity is proportional to the area of the bubble at the bubble surface and $r$ is the distance between the centres of the bubbles. Therefore, the interaction forces between bubbles which is proportional to $1/r^2$ and the product of the scattering cross sections of the two bubbles. The attractive or repulsive forces are depicted as oscillations in the volume of the two spheres which are in phase or phase opposition. The force achieves their maximum at resonance [3, 4, 7, 8].

[a] isimaciu@yahoo.com

[b] borsos.zoltan@gmail.com

The scattering-absorption phenomenon of the acoustic wave involves a cross section of interaction [7, 9].

The analogy between the electromagnetic world and the acoustic world is also advocated by the experimental and theoretical studies of other phenomena such as: the existence of the acoustic black hole [10], the acoustic Casimir effect [11-13], and the corpuscular wave duality for the acoustic wave packet [14] and for the walking droplet [15, 16].

This paper is dedicated to the study of the acoustic interaction and also of the electrostatic interaction, mentioned above, in order to highlight their analogy. The study of the two phenomena leads us to a better phenomenological understanding of the microscopic phenomena of the electromagnetic world.

# 2. The scattering and extinction cross section

## 2.1. The acoustic scattering and the extinction cross section

The acoustic cross section is analogous to the interaction cross section of an electrically charged oscillator with the field of the electromagnetic waves. Hence we will infer some correspondences between the acoustic and the electromagnetic parameters.

The interaction cross section is defined as the ratio of power (scattered and absorbed by the system) and the intensity of the incident wave [9]

$$\sigma = \frac{P}{I}. \tag{1}$$

For the acoustic waves which interact with a bubble and hence performing radial oscillations in a fluid, the power absorbed by the system is [9]

$$P = \frac{-\omega}{2\pi} \int_0^{2\pi/\omega} p_0 \left(1 + \varepsilon e^{i\omega t}\right) 4\pi R^2 \dot{R} dt = \frac{4\pi R_0 \left(p_0 \varepsilon\right)^2 \omega^2 \beta}{\rho \left[\left(\omega_0^2 - \omega^2\right)^2 + 4\beta^2 \omega^2\right]}. \tag{2}$$

and the intensity of the wave is [17, Ch.64]

$$I = \frac{\left(p_0 \varepsilon\right)^2}{\rho u}. \tag{3}$$

Substituting Eq. (2) and Eq. (3) into Eq. (1), result

$$\sigma = \frac{8\pi R_0 \omega^2 \beta u}{\left[\left(\omega_0^2 - \omega^2\right)^2 + 4\beta^2 \omega^2\right]}. \tag{4}$$

In the above relationships, we have the following notations:

$$R(t) = R_0 \left[1 + x(t)\right], \tag{5}$$

with $R_0$ the equilibrium bubble radius; the amplitude of the incident pressure wave, the forcing wave,

$$p(t) = (p_0 \varepsilon) e^{i\omega t}, \varepsilon << 1, \tag{6}$$

with $p_0$ the unperturbed fluid pressure; radial damping constant $\beta$ which is comprised of the viscous component $\beta_\mu$, the thermal component $\beta_{th}$ and the acoustic re-radiated (scattered) component $\beta_{ac}$

$$\beta = \beta_\mu + \beta_{th} + \beta_{ac}. \tag{7}$$

with

$$\beta_\mu = \frac{2\mu}{\rho R_0^2}, \beta_{th} = \frac{2\mu_{th}}{\rho R_0^2}, \beta_{ac} = \frac{\omega^2 R_0}{2u} \tag{8}$$

and the natural angular frequency of a radial oscillator

$$\omega_0 = \left[ 3\gamma \left( \frac{p_0}{\rho R_0^2} + \frac{2\sigma}{\rho R_0^3} \right) - \frac{2\sigma}{\rho R_0^3} \right]^{1/2} = \left( \frac{p_{eff}}{\rho R_0^2} \right)^{1/2}. \tag{9}$$

Replacing Eq. (7) in Eq. (4), it result in the shape of the scattering cross section and the absorption cross section

$$\sigma_e = \frac{4\pi R_0 \omega^2 u \left( \beta_\mu + \beta_{th} + \beta_{ac} \right)}{\left[ \left( \omega_0^2 - \omega^2 \right)^2 + 4\beta^2 \omega^2 \right]} = \frac{4\pi R_0 \beta_{ac} \omega^2 u}{\left[ \left( \omega_0^2 - \omega^2 \right)^2 + 4\beta^2 \omega^2 \right]} + \frac{2\pi R_0^2 \left( \beta_\mu + \beta_{th} \right) \omega^2 u}{\left[ \left( \omega_0^2 - \omega^2 \right)^2 + 4\beta^2 \omega^2 \right]} = \sigma_{ac} + \sigma_a. \tag{10}$$

Michael A. Ainslie and Timothy G. Leighton [7] distinguished three types of the cross sections for the interaction similar to the interaction of electromagnetic waves (light) with a physical system, that maybe a free electron or bound electron: the scattering cross section (re-emission, $\sigma_s = \sigma_{ac}$), the absorption cross section ($\sigma_a$) and total cross section (extinction, $\sigma_e$). Between all these sections there is relationship $\sigma_e = \sigma_s + \sigma_a$.

Comparing the expressions of the cross sections displayed in this two papers [7, 9] result a difference, observed by Ainslie and Leighton (the correct relation for time-averaged wave intensity is $I = (p_0 \varepsilon)^2 / (2\rho u)$. That is, the correct relation for (4) is

$$\sigma = \frac{8\pi R_0 \beta \omega^2 u}{\left[ \left( \omega_0^2 - \omega^2 \right)^2 + 4\beta^2 \omega^2 \right]} \tag{11}$$

and the scattering cross section is

$$\sigma_{ac} = \sigma_s = \frac{8\pi R_0 \omega^2 \beta_{ac} u}{\left[ \left( \omega_0^2 - \omega^2 \right)^2 + 4\beta^2 \omega^2 \right]} = \frac{4\pi R_0^2 \omega^4}{\left[ \left( \omega_0^2 - \omega^2 \right)^2 + 4\beta^2 \omega^2 \right]}. \tag{12}$$

## 2.2. The electrostatic cross section

The electrostatic cross section, for a particle with electric charge $Q$ and mass $m$, oscillating in electromagnetic waves is of the form [18]

$$\sigma_e = \frac{4\pi R_e \omega^2 \Gamma_t c}{\left[\left(\omega_{0e}^2 - \omega^2\right)^2 + \Gamma_t^2 \omega^2\right]}. \tag{13}$$

with $R_e = Q^2/\left(4\pi\varepsilon_0 mc^2\right) = e^2/\left(mc^2\right)$ the electrostatic radius of the particle, $\Gamma_t = \Gamma' + \omega^2\Gamma/\omega_{0e}^2$ the total decay constant. In previous relationships, $\Gamma'$ is the absorptive width, $\Gamma/\omega_{0e}^2 = 2e^2/\left(3mc^3\right) = 2R_e/\left(3c\right)$ is the radiative decay constant, $\omega_{0e}$ is the natural angular frequency and $\omega$ the angular frequency of the electromagnetic waves.

The electromagnetic scattering cross section is

$$\sigma_{se} = \frac{2}{3}\frac{4\pi R_e^2 \omega^4}{\left[\left(\omega_{0e}^2 - \omega^2\right)^2 + \Gamma_t^2 \omega^2\right]}. \tag{14}$$

By comparing the two cross sections on can establish the following correspondences:

$$R_e \leftrightarrow R_0, \Gamma_t \leftrightarrow 2\beta, c \leftrightarrow u, \omega_{0e} \leftrightarrow \omega_0. \tag{15}$$

The difference between (12) and (14) comes from the fact that the two oscillations are different, that of the bubble is radial (the centre of mass remains fixed) and that of the particle one corresponds to an oscillation of the center of mass of the oscillator.

If the particle is free, i.e. $\omega_{0e} \to 0$ or $\omega \to \infty$ then the cross section (14) becomes the Thomson cross section [18, Subch. 14.7], which is independent of the angular frequency of the incident radiation

$$\sigma_T = \frac{2}{3}\left(4\pi R_e^2\right). \tag{16}$$

The cross section for the acoustic scattering resembles the Thomson cross section in the limit cases, $\omega_0 \to 0$ or $\omega \to \infty$ and at resonance.
In the limit cases, the acoustic scattering cross section is

$$\sigma_{Bac} = \sigma_{Bs} = \lim_{\omega_0 \to 0} \sigma_s = \lim_{\omega \to \infty} \sigma_s = 4\pi R_0^2. \tag{17}$$

At speed resonance, $\omega_{rez} = \omega_0$ [19, Ch. 23], the cross section for the acoustic scattering (12) becomes

$$\sigma_{0ac} = \sigma_{0s} = 4\pi R_0^2 \left(\frac{\rho u^2}{p_{eff}}\right). \tag{18}$$

Analogues to the Thomson section, they do not depend on the angular frequency.

# 3. The acoustic force of electrostatic type

## 3.1. The secondary Bjerknes forces

The secondary Bjerknes force, the acoustic force, has been studied in several papers [1-4, 6, 9]. We consider the simpler case study of de T. Bărbat, N. Ashgriz and C-S. Hi Liu.

The expression of acoustic force for two bubbles with different radii is

$$F_B\left(\omega, r\right) = -\frac{2\pi R_{01} R_{02}}{r^2} \frac{A^2 \omega^2 \cos\varphi}{\rho\left[\left(\omega^2 - \omega_{01}^2\right)^2 + 4\beta_1^2\omega^2\right]^{1/2}\left[\left(\omega^2 - \omega_{02}^2\right)^2 + 4\beta_2^2\omega^2\right]^{1/2}}, \tag{19}$$

with $A = \varepsilon p_0$, the amplitude of the incident wave pressure.

If the bubbles are identical, $R_{01} = R_{02} = R_0$, then Eq. (19) becomes:

$$F_B(\omega, r) = \frac{-2\pi R_0^2}{r^2} \frac{p_0^2 \varepsilon^2 \omega^2}{\rho\left[\left(\omega^2 - \omega_0^2\right)^2 + 4\beta^2\omega^2\right]}, \tag{20}$$

The acoustic force is dependent on the angular frequency at the limit $\omega_0 \to 0$,

$$F_{B0}(r) = \lim_{\omega_0 \to 0} F_B(\omega, r) \cong \frac{-2\pi R_0^2}{r^2} \frac{p_0^2 \varepsilon^2}{\rho \omega^2}, \tag{21}$$

In the limit case $\omega \to \infty$, the acoustic force is zero

$$F_{B\infty}(r) = \lim_{\omega \to \infty} F_B(\omega, r) = 0. \tag{22}$$

The acoustic force is independent of angular frequency at resonance of oscillation speeds [19, Ch. 23],

$$F_B(\omega_0, r) = \lim_{\omega \to \omega_0} F_B(\omega, r) \cong \frac{-2\pi u^2 \left(p_0 \varepsilon\right)^2}{r^2 \rho \omega_0^4} \cong \frac{-2\pi R_0^4}{r^2}\left(\rho u^2\right)\left(\frac{p_0 \varepsilon}{p_{eff}}\right)^2, \tag{23a}$$

If we write the expression of force (23a) as a function of the resonance cross section (18), we obtain

$$F_B(\omega_0, r) \cong \frac{-\sigma_{0s}^2}{8\pi r^2} \frac{\left(p_0 \varepsilon\right)^2}{\rho u^2}, \tag{23b}$$

Prior studies made by various authors have highlighted the fact that for identical bubbles, $R_{01} = R_{02} = R_0$, the acoustic force is repulsive only at resonance, $R_{0r} \cong \left(p_{eff}/\rho\omega_0\right)^{1/2}$ [20-23]. Another case where the acoustic force is independent of angular frequency is that one when the bubbles interact with the acoustic background. We study this case in section 4.

## 3.2. The electrostatic force

The electrostatic force [18, Ch. 1] between two charges, the Coulomb force, is

$$F_C(r) = \frac{Q_1 Q_2}{4\pi\varepsilon_0 r^2}, \tag{24}$$

Since the charge is quantified, $Q = N\left|q_e\right|$, we can express the force according to the electron charge, $q_e$,

$$F_C(r) = N_1 N_2 \frac{q_e^2}{4\pi\varepsilon_0 r^2} = N_1 N_2 \frac{e^2}{r^2}. \tag{25}$$

In Eq. (25),

$$F_{Ce}(r) = \frac{q_e^2}{4\pi\varepsilon_0 r^2} = \frac{e^2}{r^2} \tag{26}$$

is the Coulomb force between two electrons in the vacuum. In Eq. (26), $e$ is the electrostatic charge of electron in electrostatic units. This restricts us to focus our analysis to the interaction of two identical bubbles.

### 3.3. The electrostatic force as function on the Thomson cross section

The electrostatic force between two electrons can be expressed according to the Thomson cross section, $\sigma_T = (8\pi/3)\left(e^2/m_e c^2\right)^2$ [18, Subch. 14.7]. To obtain this expression, we calculate the average energy density of the classical electron treated as a sphere with radius

$$R_e = \frac{e^2}{m_e c^2}, \tag{27}$$

according to the relationship

$$w_e = \frac{m_e}{V_e} = \frac{3m_e c^2}{4\pi R_e^3} \tag{28a}$$

or

$$m_e = \frac{4\pi R_e^3}{3c^2} w_e . \tag{28b}$$

Replacing Eq. (28b) into Eq. (27) and taking into account Eq. (16), we obtain

$$e^2 = \frac{4\pi}{3} R_e^4 w_e = \frac{3}{16\pi}\sigma_T^2 w_e . \tag{29}$$

Therefore, the expression of the Coulomb force for the electron (26) becomes

$$F_{Ce}(r) = \frac{\sigma_T^2}{8\pi r^2}\left(\frac{3w_e}{2}\right) = \frac{\sigma_T^2}{8\pi r^2}\left(\frac{9p_e}{2}\right),\ p_e = \frac{w_e}{3} . \tag{30}$$

The expression of the Coulomb force (30) and the expression of the resonant acoustic force (24) become similar only in the case when the pressure under which bubbles interact is a constant of the system. We emphasize the system consisting of a fluid in a container and bubbles in fluid.

## 4. The force and the scattering cross section in an acoustic background

### 4.1. The acoustic forces in an acoustic background

In this section we study the assumption that the acoustic forces of electrostatic type are forces between two bubbles induced by the acoustic thermal background. This background is a compound of acoustic waves with random phase or thermal acoustic radiation at equilibrium [24, Ch. 5].

The background is assumed to be done by the thermal oscillations of the cavity containing the fluid. In our picture of the background there a lot of identical bubbles and they are at resonance with each other and with the background components.

In order to calculate the force of interaction with the background we will proceed analogous to the way we have adopted when we have expressed the secondary Bjerknes forces between two bubbles. That is, we work out the volume oscillations of a bubble under the action of infinitesimal spherical pressure caused by the acoustic wave background.

The oscillations of the bubble radius is, according to Rayleigh-Plesset relationship [3, 9],

$$R_j \ddot{R}_j + \frac{3}{2}\dot{R}_j^2 = \frac{1}{\rho}\left( p_{\mathrm{int}}(t) - \frac{2\sigma}{R_j} - \frac{4\mu}{R_j}\dot{R}_j - p_{ext}(t) \right),\ j = 1,2, \tag{31}$$

with the external pressure determined by the wave

$$p_{ext}(t) = p_0 + A\cos\omega t \tag{32}$$

and $p_0$ the hydrostatic pressure of the unperturbed fluid.

For small amplitudes, the solution of eq. (31) is [3, 9].

$$R(t) = R_0\left[1 + a\cos\left(\omega t + \varphi\right)\right] \tag{33}$$

with dimensionless amplitude and phase given relationships:

$$a = \frac{A}{\rho R_0^2\left[\left(\omega^2 - \omega_0^2\right)^2 + 4\beta^2\omega^2\right]^{1/2}}, \quad \varphi = \arctan\frac{2\beta\omega}{\left(\omega^2 - \omega_0^2\right)} \tag{34}$$

and the natural angular frequency given by Eq. (9) and the damping constant given by the Eq. (7).

The correspondence between the notations adopted by Prosperetti and Bărbat is

$$p_0\varepsilon = A, \;\; x(t) = a\cos\left(\omega t + \varphi\right). \tag{35}$$

In our approach an additional pressure produced by the radial bubble oscillation has to be added to them mentioned above:

$$p'(r,t) = \frac{\ddot{V}}{4\pi r} - \rho\frac{\rho\dot{R}^2}{2}\left(\frac{R}{r}\right)^4 \cong \frac{\ddot{V}}{4\pi r} = \frac{R}{r}\left(2\dot{R}^2 + R\ddot{R}\right) \cong \frac{R^2\ddot{R}}{r}. \tag{36}$$

Substituting Eq. (33) to Eq. (36) gives

$$p'(r,t) \cong -\frac{\rho\omega^2 R_0^3 a}{r}\cos\left(\omega t + \varphi\right). \tag{37}$$

with $r$ the distance between the centres of the bubbles.

The expression of secondary Bjerknes force is:

$$\vec{F}_B = \left\langle \vec{F}_{12} \right\rangle = -\left\langle V_2(t)\nabla p_1'(r,t) \right\rangle \;, \tag{38}$$

with

$$V_2(t) = \frac{4\pi R_{02}^3}{3}\left[1 + a_2\cos\left(\omega t + \varphi_2\right)\right]^3 \tag{39}$$

and

$$\nabla p_1'(r,t) \cong \frac{\rho\omega^2 R_{01}^3 a_1}{r^2}\cos\left(\omega t + \varphi_1\right). \tag{40}$$

Substituting Eq. (40) into Eq. (38) then it follows

$$\begin{gathered} F_B\left(r\right) = \left\langle F_{12}\right\rangle = -\frac{2\pi\rho\omega^2 R_{01}^3 R_{02}^3}{r^2}a_1 a_2\cos\varphi\,\Phi\left(a_1, a_2, \varphi = \varphi_2 - \varphi_1\right) \\ \Phi\left(a_1, a_2, \varphi\right) = 1 - \frac{a_1 a_2}{\cos\varphi} + \frac{1}{4}\left(a_1^2 + a_2^2\right) + 2a_1 a_2\cos\varphi + O\left(a_1^i a_2^j\right), i + j \geq 3 \end{gathered}, \tag{41}$$

Substituting Eq. (34) into Eq. (41), then we are lead to Eq. (19).

In order to complete our approach we will assume that the pressure is an external pressure. This will be done when we will study the interaction of a bubble with the acoustic background. Let remind that waves have a certain angular frequency and a random phase (waves corresponding to an infinitely small angular frequency interval)

$$\delta p_{ext}(t) = p_0 + (\delta A)\cos(\omega t + \theta) \tag{42}$$

In what it follows we will broach the issue of how to express the acoustic force which can generate an elementary oscillation, i.e. an oscillation with elementary amplitude. The thermal background is responsible for this action.

A wave changes the pressure of a liquid according to the relation [17, Ch.8, §63]

$$p_w = p\left(\frac{\rho_w}{\rho}\right)^{\gamma} = p\left(1+\frac{\delta\rho_w}{\rho}\right)^{\gamma} = p\left(1+\frac{\dot{q}}{u}\right)^{\gamma} \tag{43}$$

with $p = p_a = p_0 = \rho u^2$ the unperturbed fluid pressure, $\dot{q}$ the speed of the oscillation of the fluid particle and $u$ the wave speed. For some liquid, $\gamma$ is replace by the adiabatic coefficient for liquids $\gamma_f$.

Assuming the notations used for the Bjerknes force, the pressure can be expressed as,

$$p_w = p\left(1+\frac{\dot{q}}{u}\right)^{\gamma_f} \cong p_0\left(1+\gamma_f\frac{\dot{q}}{u}\right) = p_0\left[1+\gamma_f\frac{q_0\omega}{u}\cos\left(\omega t - \vec{r}\vec{k} + \theta\right)\right] \tag{44}$$

with the fluid oscillation amplitude $q_0$. We have adopted in (44) a total phase $\omega t - \vec{r}\vec{k} + \theta$ in order to average over random phase, in the same manner as that one used for the Classical Zero Point Field [25, 26]).

Collating Eq. (44) with the Eq. (32) yields

$$p_{ext} = p_w,\ A = p_0\frac{\gamma_f q_0\omega}{u} = \rho u\gamma_f q_0\omega \tag{45}$$

For the acoustic background, $\delta A$ is

$$\delta A = \frac{p_0\gamma_f\omega}{u}h(\omega,\theta)\exp\left[-i\left(\omega t - \vec{k}\vec{r} - \theta\right)\right]d^3k = \frac{p_0\gamma_f\omega}{u}q_s \tag{46}$$

with $\theta$ the random phase and $h(\omega,\theta)$ the wave amplitude and the stochastic amplitude $q_s = h(\omega,\theta)\exp\left[-i\left(\omega t - \vec{k}\vec{r} - \theta\right)\right]d^3k$.

Therefore, the square of the amplitude determined by a stochastic background is

$$\begin{aligned}\left\langle A_s^2\right\rangle = \left(\frac{p_0\gamma_f}{u}\right)^2 \Big\langle \frac{\text{Re}}{2}\iint \omega d^3k\omega' d^3k' h(\omega,\theta)h(\omega',\theta')\times \\ \exp\left[i(\vec{k}-\vec{k}')\vec{r} - i(\omega-\omega')t + i(\theta-\theta')\right]\Big\rangle.\end{aligned} \tag{47}$$

Calculating the average of the exponential function for the random phase, that is for all values, $\theta \in [0, 2\pi]$, with the same probability, we find out

$$\left\langle h(\omega,\theta)h(\omega',\theta')\exp\left[i(\vec{k}-\vec{k}')\vec{r} - i(\omega-\omega')t + i(\theta-\theta')\right]\right\rangle = h(\omega)h(\omega')\delta_{\omega\omega'}\delta\left(k-k'\right) \tag{48}$$

According to Dirac function properties, $\int_{-\infty}^{\infty}\delta(x-a)f(x)dx = f(a)$, then

$$\int_{-\infty}^{\infty}\omega d^3kh(\omega)\delta(k-k') = \omega' h(\omega'). \tag{49}$$

Substituting Eqs. (49) and (48) into Eq. (47) the mean square of the amplitude is

$$\langle A_s^2\rangle = \left(\frac{p_0\gamma_f}{u\sqrt{2}}\right)^2 \int (\omega')^2 d^3k' h^2(\omega')\delta_{\omega\omega'} = \left(\frac{p_0\gamma_f}{u\sqrt{2}}\right)^2 \int \omega^2 d^3k h^2(\omega) = \frac{2\pi\left(p_0\gamma_f\right)^2}{u^5}\int \omega^4 h^2\left(\omega\right) d\omega, \tag{50}$$

with, $\omega = ku$ and $d^3k = 4\pi k^2 dk = \left(4\pi\omega^2 d\omega\right)/u^3$ .

One can use again the same way of the Classical Zero Point Field approach to express the amplitude function $h(\omega)$, because this is part of the energy density expression of the background.

The total energy of an oscillator under the action of an elementary wave of a particular angular frequency, is

$$\langle E_o\rangle = \frac{m}{2}\langle\omega^2 q_s^2\rangle = \frac{m}{2}\left\langle \frac{\mathrm{Re}}{2}\int \omega d^3k \int \omega' d^3k' h(\omega,\theta)h(\omega',\theta') \exp\left[i(\vec{k}-\vec{k}')\vec{r} - i(\omega-\omega')t + i(\theta-\theta')\right]\right\rangle = \frac{m}{4}\int \omega^2 h^2(\omega) d^3k = \frac{\pi m}{u^3}\int \omega^4 h^2(\omega) d\omega, \tag{51}$$

Energy density of a stochastic wave is

$$w_s = n\langle E_o\rangle = \frac{\pi\left(nm\right)}{u^3}\int \omega^4 h^2(\omega) d\omega = \frac{\pi\rho}{u^3}\int \omega^4 h^2(\omega) d\omega \tag{52}$$

with $n$ the fluid volume density of the particles and $\rho = nm$ the fluid density.

From the (52), the spectral density of the electromagnetic background is

$$\rho(\omega) = \frac{dw_s}{d\omega} = \frac{\pi\rho}{u^3}\omega^4 h^2(\omega) . \tag{53}$$

This spectral density has a temperature dependence of the form

$$\rho(\omega,T) = \frac{\hbar\omega^3}{\pi^2 c^3}\frac{1}{\exp\left[\hbar\omega/\left(kT\right)\right]-1} . \tag{54}$$

and for the zero temperature of the electromagnetic background

$$\rho(\omega, T=0) = \frac{\hbar\omega^3}{2\pi^2 c^3} . \tag{55}$$

For the acoustic background, these spectral densities are:

$$\rho(\omega,T) = \frac{\hbar\omega^3}{2\pi^2 u^3}\frac{1}{\exp\left[\hbar\omega/\left(kT\right)\right]-1} , \tag{56}$$

$$\rho(\omega) = \frac{\hbar\omega^3}{4\pi^2 u^3} , \tag{57}$$

because the acoustic waves travel with speed $u$ and are they are not polarized. In fluid/liquid, waves are longitudinal or compressive, therefore an additional factor $1/2$ is present in Eq. (57).

Comparing the relations (53) and (56-57), then it results

$$h(\omega,T)=\frac{\hbar^{1/2}}{\left[2\pi^{3}\rho\omega\left(\exp\frac{\hbar\omega}{kT}-1\right)\right]^{1/2}},\ h(\omega)=\frac{\hbar^{1/2}}{\left(4\pi^{3}\rho\omega\right)^{1/2}}. \qquad (58)$$

We will use (58) in order to estimate the force determined by the acoustic stochastic background.

If the pressure of the stochastic wave is given by Eq. (46), then the amplitude of the bubble oscillation ($R(t)=R_0\left[1+\delta a(t)\right]$) vary as

$$\delta a(t)=\frac{\delta A(t)}{\rho R_0^2\left[\left(\omega^2-\omega_0^2\right)^2+4\beta^2\omega^2\right]^{1/2}}=\frac{\gamma_f p_0\omega h(\omega,\theta)\exp\left[-i\left(\omega t-\vec{k}\vec{r}-\theta-\varphi\right)\right]d^3k}{\rho u R_0^2\left[\left(\omega^2-\omega_0^2\right)^2+4\beta^2\omega^2\right]^{1/2}} \qquad (59)$$

with this relation one can express the volume

$$\delta V_2(t)=\frac{4\pi R_{02}^3}{3}\left[1+\delta a_2(t)\right]^3\cong 4\pi R_{02}^3\delta a_2(t)=$$
$$\frac{4\pi\gamma_f p_0 R_{02}\omega h(\omega,\theta)\exp\left[-i\left(\omega t-\vec{k}\vec{r}-\theta-\varphi_2\right)\right]d^3k}{\rho u\left[\left(\omega^2-\omega_{02}^2\right)^2+4\beta_2^2\omega^2\right]^{1/2}} \qquad (60)$$

and the pressure

$$\delta p_1'(r,t)\cong\frac{-\rho\omega'^2R_{01}^3\delta a_1(t)}{r}=\frac{-\omega'^2R_{01}}{r}\frac{\gamma_f p_0\omega' h(\omega',\theta')\exp\left[-i\left(\omega' t-\vec{k}'\vec{r}-\theta-\varphi_1\right)\right]d^3k'}{u\left[\left(\omega'^2-\omega_{01}^2\right)^2+4\beta_1^2\omega'^2\right]^{1/2}}. \qquad (61)$$

Then Eq. (38), with $\delta V_2(t)$ and $\delta p_1'(r,t)$, expressed above, becomes

$$\delta\vec{F}_B=-\delta V_2(t)\nabla\left[\delta p_1'(r,t)\right], \qquad (62)$$

We will assume that the origin of the reference system is in the centre of the bubble 1 and the second bubble has the position vector $\vec{r}$. Substituting Eq. (60) and Eq. (61) in Eq. (62), it leads to

$$\delta F_B(r)=\frac{-4\pi R_{01}R_{02}\left(p_0\gamma_f\right)^2}{r^2\rho u^2}\frac{\left\{\omega h(\omega,\theta)\exp\left[-i\left(\omega t-\vec{k}\vec{r}-\theta-\varphi_2\right)\right]d^3k\right\}}{\left[\left(\omega^2-\omega_{02}^2\right)^2+4\beta_2'^2\omega^2\right]^{1/2}}\times$$
$$\frac{\left\{\omega'^3h(\omega',\theta')\exp\left[-i\left(\omega' t-\vec{k}'\vec{r}-\theta-\varphi_1\right)\right]d^3k'\right\}}{\left[\left(\omega'^2-\omega_{01}^2\right)^2+4\beta_1'^2\omega'^2\right]^{1/2}}. \qquad (63)$$

For the case when the bubbles are identical, Eq. (63) becomes

$$\delta F_B(r)=\frac{-4\pi R_0^2\left(p_0\gamma_f\right)^2}{r^2\rho u^2}\frac{\left\{\omega h(\omega,\theta)\exp\left[-i\left(\omega t-\vec{k}\vec{r}-\theta-\varphi\right)\right]d^3k\right\}}{\left[\left(\omega^2-\omega_0^2\right)^2+4\beta^2\omega^2\right]^{1/2}}\times$$
$$\frac{\left\{\omega'^3 h(\omega',\theta')\exp\left[-i\left(\omega' t-\vec{k}'\vec{r}-\theta-\varphi\right)\right]d^3k'\right\}}{\left[\left(\omega'^2-\omega_0^2\right)^2+4\beta'^2\omega'^2\right]^{1/2}}. \tag{64}$$

By integrating and performing the average for random phase in Eq. (64), the secondary Bjerknes force has the form

$$F_B(r)=\frac{-4\pi R_0^2\left(p_0\gamma_f\right)^2}{r^2\rho u^2}\times$$
$$\frac{\mathrm{Re}}{2}\left\langle\iint\frac{\left\{\omega h(\omega,\theta)\exp\left[-i\left(\omega t-\vec{k}\vec{r}-\theta-\varphi\right)\right]d^3k\right\}}{\left[\left(\omega^2-\omega_0^2\right)^2+4\beta^2\omega^2\right]^{1/2}}\times\right.$$
$$\left.\frac{\left\{\omega'^3 h(\omega',\theta')\exp\left[i\left(\omega' t-\vec{k}'\vec{r}-\theta-\varphi\right)\right]d^3k'\right\}}{\left[\left(\omega'^2-\omega_0^2\right)^2+4\beta'^2\omega'^2\right]^{1/2}}\right\rangle= \tag{65}$$
$$\frac{-2\pi R_0^2\left(p_0\gamma_f\right)^2}{r^2\rho u^2}\int_{k_m}^{k_M}\frac{\omega^4h^2(\omega)d^3k}{\left[\left(\omega^2-\omega_0^2\right)^2+4\beta^2\omega^2\right]}=$$
$$\frac{-8\pi^2 R_0^2\left(p_0\gamma_f\right)^2}{r^2\rho u^5}\int_{\omega_m}^{\omega_M}\frac{h^2(\omega)\omega^6d\omega}{\left[\left(\omega^2-\omega_0^2\right)^2+4\beta^2\omega^2\right]}.$$

The integration limits $\omega_M=2\pi u/\lambda_m=\pi u/a\cong\omega_D$ and $\omega_m=2\pi u/\lambda_M=\pi u/L$ correspond to the minimum wavelength, $\lambda_m=2a$ and the maximum wavelength $\lambda_M=2L$. The maximum angular frequency is approximate the Debye angular frequency, $\omega_D=(u/a)\left(6\pi^2\right)^{1/3}$ [24, Ch. 5].

In our above relationships, $a$ is the average distance between the fluid/liquide particles and $L$ is the linear dimension of the container that delimits the fluid/liquid.

Substituting Eq. (58), with $h^2(\omega)\rightarrow h^2(\omega,T)$, into Eq. (65), we finally fiind:

$$F_{BT}(r)=\frac{-4\left(p_0\gamma_f\right)^2R_0^2\hbar}{\pi\rho^2u^5r^2}\int_{\omega_m}^{\omega_M}\frac{\omega^5d\omega}{\left(\exp\frac{\hbar\omega}{kT}-1\right)\left[\left(\omega^2-\omega_0^2\right)^2+4\beta^2\omega^2\right]},$$
$$F_{BZ}(r)=\frac{-2\left(p_0\gamma_f\right)^2R_0^2\hbar}{\pi\rho^2u^5r^2}\int_{\omega_m}^{\omega_M}\frac{\omega^5d\omega}{\left[\left(\omega^2-\omega_0^2\right)^2+4\beta^2\omega^2\right]}. \tag{66}$$

## 4.2. The acoustic cross section in an acoustic background

We are entitled to make the assumption that the acoustic cross section for the radial movement becomes a cross section of the Thomson type. This can be shown if one calculates the average cross section of all frequencies of the acoustic background corresponding to the acoustic radiation of the container in which the bubble is:

$$\sigma_{aT} = \frac{\int_{\omega_m}^{\omega_M} \sigma(\omega)\rho(\omega,T)d\omega}{\int_{\omega_m}^{\omega_M} \rho(\omega,T)d\omega} = \frac{\int_{\omega_m}^{\omega_M} \left[\frac{8\pi R_0 \beta \omega^2 u}{\left[\left(\omega_0^2-\omega^2\right)^2+4\beta^2\omega^2\right]}\right]\left[\frac{\hbar\omega^3}{2\pi^2u^3}\frac{1}{\exp\left[\hbar\omega/(kT)\right]-1}\right]d\omega}{\int_{\omega_m}^{\omega_M}\left[\frac{\hbar\omega^3}{2\pi^2u^3}\frac{1}{\exp\left[\hbar\omega/(kT)\right]-1}\right]d\omega} =$$

$$\frac{8\pi u R_0 \int_{\omega_m}^{\omega_M}\frac{\beta\omega^5 d\omega}{\left[\exp\left(\hbar\omega/(kT)\right)-1\right]\left[\left(\omega_0^2-\omega^2\right)^2+4\beta^2\omega^2\right]}}{\int_{\omega_m}^{\omega_M}\frac{\omega^3 d\omega}{\exp\left[\hbar\omega/(kT)\right]-1}} \tag{67a}$$

and

$$\sigma_{az} = \frac{\int_{\omega_m}^{\omega_M} \sigma(\omega)\rho(\omega)d\omega}{\int_{\omega_m}^{\omega_M} \rho(\omega)d\omega} = \frac{\int_{\omega_m}^{\omega_M} \left[\frac{8\pi R_0 \beta u \omega^2}{\left[\left(\omega_0^2-\omega^2\right)^2+4\beta^2\omega^2\right]}\right]\left(\frac{\hbar\omega^3}{4\pi^2u^3}\right)d\omega}{\int_{\omega_m}^{\omega_M}\left(\frac{\hbar\omega^3}{4\pi^2u^3}\right)d\omega} =$$

$$\frac{8\pi u R_0 \int_{\omega_m}^{\omega_M}\frac{\beta\omega^5 d\omega}{\left[\left(\omega_0^2-\omega^2\right)^2+4\beta^2\omega^2\right]}}{\int_{\omega_m}^{\omega_M}\omega^3 d\omega}. \tag{67b}$$

We remind that the thermal background is the acoustic radiation of the container which contains the bubbles.

The same result is reached when in the definition (1) of the cross section it is inserted the power and the intensity calculated by the averaged random-phase according to the section 4.1.

# 5. The force and the acoustic cross section in the background: analytical estimation

## 5.1. The acoustic force of the electrostatic type and the acoustic charge

In order to estimate analytically the expression (66) of the force, we calculate the integral using the saddle-point method [27; 19, Ch. 23]. In Eq. (66), the integral has a maximum for $\omega \cong \omega_0$. If we make the change of variables $\omega - \omega_0 = y$, the integral becomes

$$I_B = \frac{\omega_0^5}{4\omega_0^2\left(\exp\frac{\hbar\omega_0}{kT}-1\right)}\int_{y_m}^{y_M}\frac{dy}{y^2+\beta_0^2} =$$

$$\frac{\omega_0^3}{4\beta_0\left(\exp\frac{\hbar\omega_0}{kT}-1\right)}\left[\arctan\frac{\omega_M-\omega_0}{\beta_0}-\arctan\frac{\omega_m-\omega_0}{\beta_0}\right], \tag{69}$$

with

$$\beta_0 = \beta(\omega_0) = 2\frac{\mu+\mu_{th}}{\rho R_0^2} + \frac{\omega_0^2 R_0}{2u} = \frac{\omega_0^2 R_0}{2u}\left(1+\frac{4\mu u}{\rho R_0^3\omega_0^2}+\frac{2u\beta_{th0}}{R_0\omega_0^2}\right).$$

If $\omega_M \gg \omega_0, \omega_m \ll \omega_0$, we can approximate, $\omega_M - \omega_0 \cong \omega_M = 2\pi u/\lambda_m = \pi u/a \to \infty$, $\omega_m - \omega_0 \cong -\omega_0$. With these approximations, Eq. (68) becomes

$$I_B \cong \frac{\omega_0^3}{4\beta_0\left(\exp\frac{\hbar\omega_0}{kT}-1\right)}\left(\frac{\pi}{2}+\arctan\frac{\omega_0}{\beta_0}\right) \cong$$

$$\frac{\pi u\omega_0\left(1+\frac{2}{\pi}\arctan\frac{2u}{R_0\omega_0}\right)}{4R_0\left(\exp\frac{\hbar\omega_0}{kT}-1\right)\left(1+\frac{4\mu u}{\rho R_0^3\omega_0^2}+\frac{2u\beta_{th0}}{R_0\omega_0^2}\right)} \cong \frac{\pi u\omega_0}{2R_0\left(\exp\frac{\hbar\omega_0}{kT}-1\right)\left(1+\frac{4\mu u}{\rho R_0^3\omega_0^2}+\frac{2u\beta_{th0}}{R_0\omega_0^2}\right)}. \tag{70}$$

Substituting Eq. (70) into Eq. (66), with $\gamma_f p_0 = \rho u^2$ [10]), it follows

$$F_B(R_0,\omega_0,T,r) \cong \frac{R_0\hbar\omega_0}{\left(\exp\frac{\hbar\omega_0}{kT}-1\right)\left(1+\frac{4\mu u}{\rho R_0^3\omega_0^2}+\frac{2u\beta_{th0}}{R_0\omega_0^2}\right)r^2}. \tag{71a}$$

For the acoustic CZPF background, replacing in expression (65) the square of the stochastic amplitude, $h_a^2(\omega) = \hbar/\left(4\pi^3\rho\omega\right)$, results

$$F_B(R_0,\omega_0,r) = \left(\frac{-\hbar u}{r^2}\right)\frac{\omega_0 R_0}{2u\left(1+\frac{4\mu u}{\rho R_0^3\omega_0^2}+\frac{2u\beta_{th0}}{R_0\omega_0^2}\right)}. \tag{71b}$$

The expression of the acoustic force magnitude given by Eq. (71) can be put under the form

$$F_a = \frac{e_a^2}{r^2}, \tag{72}$$

in order to suggest the existence of an acoustic charge (equivalent of charge in electrostatic units), $e_a$.

The square of the acoustic thermal charge is

$$e_{aT}^2 = \frac{\hbar R_0\omega_0}{\left(\exp\frac{\hbar\omega_0}{kT}-1\right)\left(1+\frac{4\mu u}{\rho R_0^3\omega_0^2}+\frac{2u\beta_{th0}}{R_0\omega_0^2}\right)} = \frac{(\hbar u)R_0\omega_0}{u\left(\exp\frac{\hbar\omega_0}{kT}-1\right)\left(1+\frac{4\mu u}{\rho R_0^3\omega_0^2}+\frac{2u\beta_{th0}}{R_0\omega_0^2}\right)}. \tag{73a}$$

and square of the acoustic zero point charge is

$$e_{aZ}^2 = \frac{\hbar R_0\omega_0}{2\left(1+\frac{4\mu u}{\rho R_0^3\omega_0^2}+\frac{2u\beta_{th0}}{R_0\omega_0^2}\right)} = \frac{(\hbar u)R_0\omega_0}{2u\left(1+\frac{4\mu u}{\rho R_0^3\omega_0^2}+\frac{2u\beta_{th0}}{R_0\omega_0^2}\right)}. \tag{73b}$$

Depending on the ratio between $\hbar\omega_0$ and $kT$, we get the following expressions of the square of the acoustic thermal charge:

$$e_a^2 = \frac{kTR_0}{\left(1+\frac{4\mu u}{\rho R_0^3\omega_0^2}+\frac{2u\beta_{th0}}{R_0\omega_0^2}\right)}, \hbar\omega_0 << kT; \tag{74}$$

$$e_{aT}^2 = \frac{R_0 \hbar \omega_0}{(e-1)\left(1+\frac{4\mu u}{\rho R_0^3 \omega_0^2}+\frac{2u\beta_{th0}}{R_0\omega_0^2}\right)}, \hbar\omega_0 \cong kT; \tag{75}$$

$$e_{aT}^2 = \frac{\hbar R_0 \omega_0}{\left(1+\frac{4\mu u}{\rho R_0^3 \omega_0^2}+\frac{2u\beta_{th0}}{R_0\omega_0^2}\right)\exp\frac{\hbar\omega_0}{kT}}, \hbar\omega_0 >> kT. \tag{76}$$

## 5.2. The acoustic cross section in the background

Now we will estimate analytically the expression (67) of the acoustic cross section in the background. To do this we calculate the integrals using the same method as in the section 5.1:

$$I_{\sigma T1} = \int_0^\infty \frac{\beta\omega^5 d\omega}{\left[\exp\left(\hbar\omega/(kT)\right)-1\right]\left[\left(\omega_0^2-\omega^2\right)^2+4\beta^2\omega^2\right]} = \frac{\beta_0\omega_0^3}{4\left[\exp\left(\hbar\omega_0/(kT)\right)-1\right]}\int_{\omega_0}^\infty \frac{-dy}{y^2+\beta_0^2} \cong \frac{\pi\omega_0^3}{4\left[\exp\left(\hbar\omega_0/(kT)\right)-1\right]} \text{ and} \tag{77a}$$

$$I_{\sigma Z1} = \int_0^{\omega_M} \frac{\beta\omega^5 d\omega}{\left[\left(\omega_0^2-\omega^2\right)^2+4\beta^2\omega^2\right]} \cong \frac{\beta_0\omega_0^3}{4}\int_{\omega_0}^\infty \frac{-dy}{y^2+\beta_0^2} \cong \frac{\pi\omega_0^3}{4}. \tag{77b}$$

The second integrals is:

$$I_{\sigma T2} = \int_{\omega_m}^{\omega_M} \frac{\omega^3 d\omega}{\exp\left[\hbar\omega/(kT)\right]-1} = \left(\frac{kT}{\hbar}\right)^4 \int_{x_m}^{x_M} \frac{x^3 dx}{\exp x - 1} \cong \left(\frac{kT}{\hbar}\right)^4 \int_0^\infty \frac{x^3 dx}{\exp x - 1} \cong \frac{\pi^4}{15}\left(\frac{kT}{\hbar}\right)^4 \text{ and} \tag{78a}$$

$$I_{\sigma Z2} = \int_{\omega_m}^{\omega_M} \omega^3 d\omega \cong \int_0^{\omega_M} \omega^3 d\omega \cong \frac{\omega_M^4}{4}. \tag{78b}$$

Replacing the two integrals (77) and (78) into Eq. (67), it results:

$$\sigma_{aT} = \frac{30R_0\omega_0^3 u}{\pi^2\left(\frac{kT}{\hbar}\right)^4\left[\exp\left(\hbar\omega_0/(kT)\right)-1\right]} = \frac{30R_0\omega_0^3 u}{\pi^2\omega_T^4\left[\exp\left(\hbar\omega_0/(kT)\right)-1\right]}, \omega_T = \frac{kT}{\hbar} \text{ and} \tag{79a}$$

$$\sigma_{aZ} = \frac{8\pi^2 R_0\omega_0^3 u}{\omega_M^4}. \tag{79b}$$

Depending on the ratio between $\hbar\omega_0$ and $kT$ , we get the following expressions of the thermal scattering cross section:

$$\sigma_{aT1} = \frac{30R_0\omega_0^2 u}{\pi^2\omega_T^3}, \hbar\omega_0 << kT; \tag{80}$$

$$\sigma_{aT2} = \frac{30R_0 u}{\pi^2(e-1)\omega_0} = \frac{30R_0^2 u}{\pi^2(e-1)}\sqrt{\frac{\rho u^2}{p_{eff}}}, \ \hbar\omega_0 \cong kT; \tag{81}$$

$$\sigma_{aT3} = \frac{30R_0\omega_0^3 u}{\pi^2\omega_T^4\exp\left(\hbar\omega_0/(kT)\right)}, \hbar\omega_0 >> kT. \tag{82}$$

The expression (79b) of the cross section corresponding to the CZPF acoustic background becomes similar to that for the thermal background (81) if $\omega_M \cong \omega_T = kT/\hbar \cong \omega_0$ .

The relationships (73) highlights the existence of a natural acoustic charge

$$e_{an}^2 = \hbar u. \tag{83}$$

According to relations (73), depending on the values of the liquid temperature $T$ and bubble radius $R_0$, the thermal and acoustic zero point charges can be lower or higher than the natural acoustic charge.

This happens similarly to the electromagnetic world, where there is a natural maximum charge of interaction, $e_{en}^2 = \hbar c$, such that $\hbar c/e^2 = e_{en}^2/e^2 \cong 137$ [18, Subch. 6.12]. This natural charge is also the maximum charge in the electromagnetic world.

## 5.3. Acoustic field

By analogy with the electrostatic field [18, Subch. 1.2], the acoustic charge, at rest, produces a static acoustic field of intensity $\vec{E}_a$:

$$\vec{E}_a = \frac{\vec{F}_a}{e_a} = \frac{e_a}{r^3}\vec{r}\,, \tag{84}$$

The energy density of this acoustic field is

$$w_a(r) = \frac{1}{8\pi}\vec{E}_a^2 = \frac{e_a^2}{8\pi r^4}. \tag{85}$$

If we consider the bubble with radius $R_a$ as a system that scatters acoustic waves, the acoustic energy density on the scattering surface is

$$w_a(R_a) = \frac{e_a^2}{8\pi R_a^4}. \tag{86}$$

By analogy to electrostatics, the equivalent mass of this system is

$$m_a = V_a\frac{w_a(R_a)}{u^2} = \frac{4\pi R_a^3 w_a(R_a)}{3u^2} = \frac{e_a^2}{6u^2 R_a}. \tag{87}$$

This mass is related to the mass corresponding to the acoustic field up to a constant. This mass is different from the virtual mass of the bubble $m_{vb} = (2/3)\pi R_0^3\rho$ [17, Ch. 1; 3]. The mass of the acoustic energy of the field generated by the bubbles is

$$m_{fa} = \frac{1}{u^2}\int_{R_a}^{L} w_a 4\pi r^2 dr = \frac{e_a^2}{2u^2 R_a} - \frac{e_a^2}{2u^2 L} \cong \frac{e_a^2}{2u^2 R_a},\ L >> R_a. \tag{88}$$

where $L$ is the radius of the container.

If in Eq. (87) we consider the last equality (88), then it follows the relationship

$$e_a^2 = 8\pi R_a^4 w_a(R_a) = \frac{\sigma_a^2}{2\pi} w_a(R_a), \tag{89}$$

with $\sigma_a = 4\pi R_a^2$. The relationship (89) is analogous to the electrostatic relationship given by Eq. (29).

From the relation $\sigma_a = 4\pi R_a^2$ and Eq. (79), one can express the radii of the system which scatter the acoustic waves, through the oscillating bubble parameters:

$$R_{aT} = \sqrt{\frac{15 R_0\omega_0^3 u}{2\pi^3\omega_T^4\left[\exp\left(\hbar\omega_0/(kT)\right)-1\right]}},\ R_{aZ} = \sqrt{\frac{2\pi R_0\omega_0^3 u}{\omega_M^4}}. \tag{90}$$

In the particular case given by the condition of Eq. (81), $\hbar\omega_0 \cong kT$ , a simple relationship follows

$$R_{aT2} = R_0 \sqrt{\frac{15}{2\pi^3 (e-1)} \sqrt{\frac{\rho u^2}{p_{eff}}}}. \tag{91}$$

# 6. Conclusions

Previous papers on the issue of the analogy between the electrostatic and the acoustic forces were based on the fact that both forces depend inverse proportional to the square of the distance between the physical systems in interaction. Also the analogy have taken account of the symmetrically dependence of the forces on the parameters of the two systems, the particles and the bubbles, $(e_1, e_2)$ and $(R_{01}, R_{02})$, and the inductive acoustic wave parameters $(\varepsilon p_0, \omega)$. For this reason, the acoustic force was considered to be similar to the gravitational force between two masses.

We showed in this paper that the acoustic force caused by the scattering process $(\beta_{ac} >> \beta_\mu + \beta_{th})$ cannot be analogous to the gravitational force because it is repulsive, $\varphi \in (\pi/2, \pi]$, and also attractive, $\varphi \in [0, \pi/2)$. This situation is similar to the electrostatic interaction.

The acoustic force given by the Eq. (71) is not quite an electrostatic type force because in the fluid the bubbles have different radii and no quantification of the radii was observed. It is necessary to study the formation and evolution of vapour bubbles (without gas) to find out the conditions in which they have various values of very close radii. In electrostatics, systems can have various electrical charges through the accumulation of elementary charges, which are the charges carried by the electron and the proton. In our paper, we studied the limited case of the interaction of two identical bubbles.

For identical bubbles, we identified the existence of an acoustic charge and an acoustic cross section. The acoustic scattering cross section, which has the definition in Eq. (1), does not depend on the amplitude of the forcing wave but depends on the properties of the bubble. The bubble is assumed to be a forced radial oscillator with damping process and with the centre of the bubble at rest. The cross section of the free electron scattering, according to Eq. (16) is independent of the angular frequency. The acoustic charge also depends on the amplitude of the forcing wave. These two parameters, the cross section and the acoustic charge, are not related of the angular frequency for angular frequency close to the natural angular frequency that is at resonance. In this case, the two parameters remain dependent on the magnitude of the forcing wave. In order to eliminate the two dependencies we had two options to consider: the two bubbles interact with the background of the thermal radiation or the two bubbles interact with the background created by other identical oscillating bubbles in the container. Adopting the first option, we have obtained an acoustic charge and a scattering cross section analogous to electrostatic ones. We can say that in this former case the acoustic interaction is analogous to the electrostatic interaction. Applying this approach of the electrostatic interaction to the electromagnetic world, i.e. for the electron, one can be obtained a good evidence for the connection between the blackbody radiation, the relativity, and the discrete charge in classical electrodynamics [28]. The latter case will be analysed in a further paper.

We appreciate that the approach would be complete when a theoretical and an experimental evidence of a magnetic type interaction for oscillating bubbles in the translational motion would be revealed. Also, when should be revealed a magnetic field type around a bubble which performs a rotation with constant angular velocity.